\documentclass[11pt]{article}
\usepackage[T1]{fontenc}
\usepackage{lmodern}
\usepackage[hmargin=1in,vmargin=0.92in]{geometry}
\usepackage{microtype}
\usepackage{amsmath,amssymb,amsthm}
\usepackage{xcolor}
\usepackage[colorlinks=true,allcolors=blue!55!black]{hyperref}
\hypersetup{pdftitle={Separating Kolmogorov\textendash Loveland randomness from Martin-L\"of randomness},pdfauthor={Chenghua Liu, Boning Meng}}

\newtheorem{theorem}{Theorem}
\newtheorem{lemma}[theorem]{Lemma}
\newcommand{\N}{\mathbb N}

\title{Separating Kolmogorov--Loveland randomness\\
       from Martin-L\"of randomness}
\author{%
  Chenghua Liu, Boning Meng\\[0.25em]
  \small Renormalization Group\enspace\textperiodcentered\enspace
    \href{https://r13n.group/}{r13n.group}\\
  \small Correspondence:
    \href{mailto:renormalization@r13n.group}{renormalization@r13n.group}%
}
\date{}

\begin{document}
\maketitle
\vspace{-1.1em}

\begin{abstract}
Martin-L\"of randomness requires an infinite binary sequence to pass every
effective statistical test for fair coin tosses. Kolmogorov--Loveland
randomness requires every computable betting strategy to have bounded
capital, even when earlier answers guide its choice of which unread bit to
bet on next. We prove that Martin-L\"of randomness is strictly stronger. We
construct a probability measure by imposing parity equations on infinitely
many disjoint finite blocks. The systems have short descriptions relative
to their ranks, so every sequence in the support fails one fixed
Martin-L\"of test. Almost every sequence under this measure is
Kolmogorov--Loveland random. To prove this, we choose each block from a
list in which every position occurs in only a small fraction of the
candidates. For each betting strategy, after a normalization that preserves
success, this sparsity ensures that the sum of its wager fractions on bits
forced by the equations and earlier answers is almost surely finite.
\end{abstract}

\section{Introduction}

For independent fair coin tosses, the strong law of large numbers says
that the frequency of ones converges to \(1/2\) with probability one.
Martin-L\"of randomness gives a condition on an individual binary
sequence that guarantees it obeys this law and many others. A
Martin-L\"of random sequence must pass every effective statistical
test~\cite{MartinLof}. The \(k\)th level of such a test is a set of
fair-coin measure at most \(2^{-k}\), and a sequence fails the test
precisely when it lies in every level. Martin-L\"of randomness is a
standard notion in the theory of algorithmic randomness, used both to
study what can be computed from a random sequence~\cite{GreenbergMillerNies}
and to identify individual sequences for which ergodic theorems
hold~\cite{BienvenuEtAl}.

In Kolmogorov--Loveland randomness, a computable betting strategy
chooses, on the basis of the bits already seen, both an unread position
and how much to wager on its value~\cite{MSU}. It may inspect a later
bit without wagering and use the answer to bet on an earlier bit. A
sequence is Kolmogorov--Loveland random if every such strategy has
bounded capital on it. This notion has been used to study how much
power a betting strategy gains from choosing the order in which it
reads the bits~\cite{KastermansLempp}.

Each such strategy gives rise to a Martin-L\"of test that detects every
sequence on which its capital is unbounded, so every Martin-L\"of random
sequence is Kolmogorov--Loveland random. Strategies that read from left
to right are known to miss some failures of Martin-L\"of
randomness~\cite{Schnorr}. Does the freedom to choose the order of play
allow a computable betting strategy to succeed on any given sequence
that fails a Martin-L\"of test (see~\cite[Question~6.2]{MillerNies}
and~\cite{PetrovicEncoding})?

Merkle et al.~\cite{MMNRS} proved that whenever a Kolmogorov--Loveland
random sequence is computably split into two infinite subsequences, at
least one of them is Martin-L\"of random. Petrovi\'c~\cite{PetrovicOpen} found effectively open sets of arbitrarily
small fair-coin measure, each containing a sequence on which every
strategy stays within a bound depending only on the strategy. That
sequence may depend on the set. We use congruence conditions on
finitely many bits, as in Petrovi\'c's proof, and impose them on
infinitely many disjoint blocks. The resulting measure is supported on
sequences that fail one fixed Martin-L\"of test, and almost every
sequence under it is Kolmogorov--Loveland random. This gives the strict
inclusion in the following theorem.

\begin{theorem}
\label{thm:main}
The Martin-L\"of random sequences form a proper subset of the
Kolmogorov--Loveland random sequences. More precisely, there exist a
probability measure \(\rho\) on infinite binary sequences and a
Martin-L\"of test \((U_m)_{m\ge1}\) such that
\[
  \operatorname{supp}(\rho)\subseteq\bigcap_{m\ge1}U_m,
  \qquad\text{and}\qquad
  \rho\text{-almost every sequence is Kolmogorov--Loveland random.}
\]
\end{theorem}

Results on extracting Martin-L\"of random subsequences have used
Kolmogorov--Loveland randomness as a hypothesis~\cite{MMNRS}, while
effective ergodic theorems have used Martin-L\"of
randomness~\cite{BienvenuEtAl}. Theorem~\ref{thm:main} shows that these
results rest on distinct randomness assumptions, with
Kolmogorov--Loveland randomness imposing the weaker condition.

For every sequence \(X\) in the support of \(\rho\), membership in each
test level \(U_k\) is witnessed by a finite prefix of \(X\). A parity
equation illustrates the difference between finding such a witness and
placing a bet. Take three fair bits conditioned on \(x_a+x_b+x_c=0\)
over \(\mathbb F_2\). Once all three values are available, the test can
check the equation. A bettor that knows the triple can read any two
bits without wagering. Their values are independent and fair, but they
determine the third, on which the bettor can double its capital. To
exploit the equation this way it must identify the triple while one
bit is still unread.

To construct \(\rho\), we choose a uniform solution of a system of
independent parity equations on each of infinitely many disjoint finite
blocks, and a fair bit at each remaining position, independently.
The test can include many candidate systems without having to identify
the chosen one. The solution set of a rank-\(r\) system has fair-coin
measure \(2^{-r}\). There are \(2^d\) programs of length \(d\) to
describe both the equations and their positions. Among the systems
they describe, those of rank at least \(r\) thus have solution sets
of combined measure at most \(2^{d-r}\).
At level \(k\) the test sets a rank threshold for each \(d\) so that
these bounds sum to at most \(2^{-k}\). Section~\ref{sec:construction}
shows that each chosen system meets its rank threshold. By
\eqref{eq:test-criterion} of Section~\ref{sec:test}, every sequence in
the support of \(\rho\) therefore belongs to every \(U_k\).

A betting strategy can interleave its queries across blocks and use
answers from one block to choose its wagers on another. Within each
block, the equations and previous answers determine the next queried
bit exactly when deleting its column from the matrix of unread columns
lowers the rank. We call this a \emph{rank drop}. At every other query
the answer is conditionally fair. We normalize each strategy, preserving
success, so that the fractions of capital it wagers have a bounded sum
of squares. On the fair queries, the signed returns then have an almost
surely convergent sum that bounds the cumulative change in log capital
on those queries from above. If \(a_\ell\) is
the fraction wagered at the \(\ell\)th rank drop, even winning all
these bets multiplies capital by at most
\(\prod_\ell(1+a_\ell)\leq\exp(\sum_\ell a_\ell)\). It therefore
suffices to make \(\sum_\ell a_\ell\) finite \(\rho\)-almost surely.

At stage \(k\) we choose a finite set \(G\) of positions, the \emph{pool},
from which we select the block \(S\) for level \(U_k\). Under fair-coin
answers, each of the first \(k\) strategies is likely either to read
all of \(G\) or to leave a fixed fraction unread. In the latter case,
for most candidate pairs \((S,B)\) the unread columns of the matrix
\(B\) retain full rank throughout the run. A strategy that reads all of
\(G\) necessarily encounters rank drops. Since each pool position occurs
in only a small fraction of the candidate supports, the bound on squared
wager fractions makes the total fraction wagered on \(S\) small on
average over the candidates. Section~\ref{sec:construction} uses the
estimates of Section~\ref{sec:preparing} to choose one pair for all
\(k\) strategies. Section~\ref{sec:capital} sums the contributions
of the chosen blocks under \(\rho\).

\section{Preliminaries}
\label{sec:preliminaries}

Let \(\N=\{0,1,\ldots\}\). Write \(2^{<\N}\) for the set of finite
binary strings and \(2^\N\) for the space of infinite binary sequences.
Let \(\lambda\) be fair-coin measure on \(2^\N\). The \emph{cylinder} \([v]\) consists of
all sequences extending \(v\in 2^{<\N}\).
A \emph{Martin-L\"of test} is a sequence \((U_m)_{m\ge1}\) of uniformly
effectively open sets with \(\lambda(U_m)\le 2^{-m}\). Given \(m\), a
single algorithm enumerates cylinders whose union is \(U_m\). A sequence
fails the test if it belongs to every \(U_m\), and is
\emph{Martin-L\"of random} if it fails no such test.

A \emph{scan rule} is a total computable map \(\pi:2^{<\N}\to\N\)
that chooses distinct positions along every chain of prefixes. On
input \(X\), start with the empty transcript \(v_0=\epsilon\). After
\(t\) queries, the next position is \(\pi(v_t)\), its bit is
\(b_t=X(\pi(v_t))\), and the new transcript is \(v_{t+1}=v_tb_t\).
A computable non-monotonic betting strategy consists of a scan rule
\(\pi\) and a uniformly computable nonnegative capital function \(d\)
satisfying
\[
 d(\epsilon)=1,\qquad 2d(v)=d(v0)+d(v1).
\]
When \(d(v)>0\), write \(h(v)\) for the signed wager fraction, so that
an answer \(b\) changes the capital from \(d(v)\) to
\[
 d(vb)=d(v)\bigl(1+(2b-1)h(v)\bigr).
\]
The strategy wagers a fraction \(|h(v)|\) of its current capital,
betting on \(1\) if \(h(v)>0\) and on \(0\) if \(h(v)<0\).
The strategy \emph{succeeds} on \(X\) if \(\sup_t d(v_t)=\infty\).
A sequence is \emph{Kolmogorov--Loveland random} if no computable
strategy succeeds on it. Throughout, we use the equivalent formulation
of Kolmogorov--Loveland randomness in terms of total computable betting
strategies~\cite[discussion following Definition~6.1]{MillerNies}.

Fix a scan rule \(\pi\) and a finite transcript \(v=b_0\cdots b_{n-1}\).
Write \(v_t\) for the prefix of \(v\) of length \(t\). The pairs
\((\pi(v_t),b_t)\), \(t<n\), give a \emph{finite assignment} of values
to the queried positions. In transcript space, \([v]\) is the cylinder
of transcripts extending \(v\). Its preimage in input space is
\[
 C_\pi(v)=\{X:X(\pi(v_t))=b_t\text{ for all }t<n\}.
\]
Let \(\kappa\) be the law of the transcript when the input has law
\(\lambda\). Since the \(|v|\) queried positions are distinct,
\[
 \kappa([v])=\lambda(C_\pi(v))=2^{-|v|}.
\]
In particular, every scan rule induces the same fair-coin law
\(\kappa\) on transcripts. For a fixed strategy, the inputs on which its
capital ever exceeds \(2^m\) form an effectively open set, uniformly in
\(m\). By the martingale maximal inequality this set has measure at most
\(2^{-m}\). These sets form a Martin-L\"of test that covers every input
on which the strategy succeeds. Hence every Martin-L\"of random
sequence is Kolmogorov--Loveland random.

\section{Affine constraints and randomness tests}
\label{sec:test}

On a finite set \(S\subset\N\) of positions, consider a system
\(Bx=y\) of \(r\) independent parity equations over \(\mathbb F_2\).
Here \(x\in\mathbb F_2^S\) is an assignment of bits to the positions
in \(S\), and \(x_U\) denotes its restriction to \(U\subseteq S\).
Each row of \(B\) gives one equation, the columns are indexed by \(S\),
and the right-hand side \(y\in\mathbb F_2^r\) is called the \emph{syndrome}.
The event that an input satisfies this affine constraint is
\[
 E(S,B,y)=\{X\in 2^{\N}:B(X|_S)=y\}.
\]
There are \(2^{|S|-r}\) solutions on \(S\), so
\(\lambda(E(S,B,y))=2^{-r}\).

\paragraph{Rank drops and forced answers.}
Under \(\lambda\) conditioned on \(E(S,B,y)\), the assignment on \(S\)
is uniform over the solutions of \(Bx=y\). Let \(U\subseteq S\) be the
unread positions in \(S\). Given compatible answers on \(S\setminus U\),
the assignment on \(U\) is uniform over the solutions of \(B_U x_U=z\),
where \(B_U\) consists of the columns indexed by \(U\) and \(z\) is the
syndrome after subtracting the contribution of the observed bits.
Consider a query at \(q\in U\). Write \(c_q\) for its column and
\(D=B_{U\setminus\{q\}}\). The completions giving answer \(b\)
correspond to the solutions of \(D x_{U\setminus\{q\}}=z-bc_q\).
If \(c_q\) belongs to the column span
of \(D\), the systems for \(b=0,1\) have equally many solutions,
so the answer is a fair bit. Otherwise exactly one of \(z,z-c_q\) is in that
span and the answer is determined. Thus the constraint and the earlier
answers force the next bit precisely when deleting its column causes a
\emph{rank drop}, that is, when
\(\operatorname{rank} B_{U\setminus\{q\}}<\operatorname{rank} B_U\).
For a given system, this criterion depends only on \(U\) and \(q\).

Fix a transcript \(v\) of \(\pi\), and let \(U\) be the positions in
\(S\) still unread. Under \(\lambda\), conditioning on \(C_\pi(v)\)
fixes the queried bits and leaves the unread bits independent and fair.
If \(B_U\) has rank \(r\), all \(2^r\) syndromes remain equally likely.
Thus \(\lambda\bigl(E(S,B,y)\mid C_\pi(v)\bigr)=2^{-r}
=\lambda\bigl(E(S,B,y)\bigr)\). By Bayes' rule,
\[
 \lambda\bigl(C_\pi(v)\mid E(S,B,y)\bigr)
 =\frac{\lambda\bigl(E(S,B,y)\mid C_\pi(v)\bigr)\,\lambda(C_\pi(v))}
        {\lambda\bigl(E(S,B,y)\bigr)}
 =\lambda(C_\pi(v)).
\]
We will use this identity at the transcript just before the first
rank drop, when the unread columns still have full rank.

\paragraph{One test for constraints with short descriptions.}
Fix a universal interpreter for plain binary programs. At level \(m\ge1\),
run all programs in parallel. If \(p\) outputs a valid system \((S,B,y)\)
of rank \(r\ge 2|p|+m+1\), enumerate \(E(S,B,y)\) into \(U_m\).
Validity and rank are decidable from the output, and the assignments on
\(S\) satisfying the equations give a finite union of cylinders. The
sets \(U_m\) are therefore uniformly effectively open. Each accepted
program of length \(d\) contributes at most \(2^{-2d-m-1}\) in measure.
Summing over program lengths gives
\[
 \lambda(U_m)\le\sum_{d\ge0}2^d\,2^{-2d-m-1}=2^{-m}.
\]

For any sequence of such systems \((S_j,B_j,y_j)\), write
\(E_j=E(S_j,B_j,y_j)\) and \(r_j=\operatorname{rank} B_j\).
Suppose that for each \(j\) a program \(p_j\) outputs \((S_j,B_j,y_j)\).
Then
\begin{equation}
 \bigl(\forall j\ge1,\ r_j\ge 2|p_j|+j+1\bigr)
 \quad\Longrightarrow\quad
 \bigcap_{j\ge1}E_j\subseteq\bigcap_{m\ge1}U_m .
 \label{eq:test-criterion}
\end{equation}
Indeed, the rank bound ensures that the test includes \(E_j\) in \(U_j\)
for each \(j\).

\section{Preparing the construction}
\label{sec:preparing}

By Section~\ref{sec:test}, short descriptions of the block constraints
will ensure that every sequence satisfying them fails a Martin-L\"of
test. For Kolmogorov--Loveland randomness we must bound the capital
of a strategy that can choose both where to query and how much to bet.
Section~\ref{sec:budget} bounds the sum of its squared wager fractions
while preserving success. Section~\ref{sec:pools} supplies a pool
adapted to the positions the strategies query, and
Section~\ref{sec:sparse-matrices} supplies candidate supports and
matrices within that pool. Section~\ref{sec:construction} uses the
three lemmas to choose the constraint on each block.

\subsection{Square-summable wager fractions}
\label{sec:budget}

\begin{lemma}
\label{lem:budget}
For every total computable strategy with capital function \(d\), there
is a total computable strategy with the same scan rule that succeeds
on every input on which the original strategy succeeds. Its capital is
positive and rational, with initial value \(1\). Its signed wager
fraction \(h\) is a computable rational-valued function. Along every
infinite transcript, these fractions satisfy
\begin{equation}
 \sum_{t\ge0}h(v_t)^2\le\tfrac12 .
 \label{eq:budget}
\end{equation}
\end{lemma}

\begin{proof}
First we replace \(d\) by a positive rational-valued capital function
that is unbounded wherever \(d\) is. Keeping half the initial capital
in reserve and following \(d\) with the other half gives
\(c_*(v)=(1+d(v))/2\ge1/2\). To approximate \(c_*\) while retaining
fairness, set \(c(\epsilon)=1\) and, at each transcript \(v\), compute a
rational \(a(v)\) within \(2^{-|v|-3}\) of the increment
\(c_*(v1)-c_*(v)=(d(v1)-d(v0))/4\).
Set \(c(v1)=c(v)+a(v)\) and \(c(v0)=c(v)-a(v)\), so \(c\) is fair.
Along each finite transcript the accumulated error is less than \(1/4\).
Hence \(c>1/4\) everywhere, and \(c\) is unbounded wherever \(d\) is
unbounded. The signed wager fraction is \(f(v)=a(v)/c(v)\), with
\(|f(v)|<1\) since both successor capitals are positive.

Next we scale the wager fractions to bound their sum of squares on
every transcript without losing success on any input. Let \(Q_t\)
be the sum of the squared fractions wagered so far by the strategy
with capital \(c\). We use \(Q_t\) to choose the next scaling factor.
\[
 Q_t=\sum_{s<t} f(v_s)^2,\qquad
 \alpha_t=\frac{1}{2(1+Q_t)},\qquad h(v_t)=\alpha_t f(v_t).
\]
The transcript \(v_t\) determines the computable rational \(h(v_t)\).
Since \(Q_{t+1}-Q_t=f(v_t)^2<1\),
\[
 h(v_t)^2
 =\frac{Q_{t+1}-Q_t}{4(1+Q_t)^2}
 \le \frac12\left(\frac1{1+Q_t}-\frac1{1+Q_{t+1}}\right).
\]
Summing over \(t\) gives \eqref{eq:budget} by telescoping, since \(Q_0=0\).

It remains to show that the smaller wagers preserve success. Write
\(x_t=(2b_t-1)f(v_t)\) for the signed return when the answer is \(b_t\),
and put \(M_n=\sum_{t<n}x_t\). Since \(c(v_{t+1})=c(v_t)(1+x_t)\),
the inequality \(\log(1+x)\le x-x^2/4\) for \(-1<x\le1\) gives
\begin{equation}
 M_n\ge\log c(v_n)+Q_n/4\ge-\log4+Q_n/4 .
 \label{eq:linear-lower}
\end{equation}
Here and below, \(\log\) denotes the natural logarithm. Let \(D_n\) be
the normalized strategy's capital after \(n\) queries. It starts at
\(D_0=1\) and updates by \(D_{t+1}=D_t(1+\alpha_t x_t)\), so it is
positive, rational and fair. Put \(A_n=\sum_{t<n}\alpha_t x_t\).
Since \(|\alpha_t x_t|\le1/2\) and \((\alpha_t x_t)^2=h(v_t)^2\),
the inequality \(\log(1+y)\ge y-2y^2\) for \(|y|\le1/2\) and
\eqref{eq:budget} give
\[
 \log D_n=\sum_{t<n}\log(1+\alpha_t x_t)
 \ge A_n-2\sum_{t<n}h(v_t)^2\ge A_n-1 .
\]

Fix a transcript on which \(d\) is unbounded. It suffices to show
that \(A_n\) is unbounded above. Summation by parts gives
\[
 A_n=\alpha_{n-1}M_n+
       \sum_{t=1}^{n-1}(\alpha_{t-1}-\alpha_t)M_t
       \qquad(n\ge1).
\]
The coefficients are nonnegative and sum to \(\alpha_0=1/2\).

If \(Q_n\) is bounded, then \(\alpha_{n-1}\) stays bounded away from
zero. Since \(c(v_n)\) is unbounded, \eqref{eq:linear-lower} makes
\(M_n\) unbounded above, and hence so is \(\alpha_{n-1}M_n\).
The remaining sum is at least \(-(\log4)/2\), since \(M_t\ge-\log4\).
Thus \(A_n\) is unbounded above.

If \(Q_n\to\infty\), substituting \(M_t\ge-\log4+Q_t/4\) into the
same identity and summing by parts again gives
\[
 A_n\ge-\tfrac12\log4+
        \tfrac18\sum_{t<n}\frac{Q_{t+1}-Q_t}{1+Q_t}
     \ge-\tfrac12\log4+\tfrac18\log(1+Q_n).
\]
The last inequality follows by integrating \(1/(1+u)\) over each
interval \([Q_t,Q_{t+1}]\). Its right-hand side tends to infinity,
so \(A_n\to\infty\). In both cases \(D_n\) is unbounded, as required.
\end{proof}

\subsection{Pools of query positions}
\label{sec:pools}

Fix an answer sequence \(Z\in 2^\N\) and give each scan rule \(\pi_i\)
the successive bits of this same \(Z\). The rules may query different
positions. Let \(g_{i,a}(Z)\) be the indicator that \(\pi_i\) ever
queries \(a\) in this run. We use the Borel sets of transcript space,
the smallest collection containing the cylinders and closed under
complements and countable unions.

For each rule we choose a partition \(H_i,L_i\) of transcript space
once for all stages. On \(H_i\) the rule should read the whole pool,
and on \(L_i\) it should leave a fixed fraction unread. The lemma
bounds the fair-coin measure of the exceptions in each part. For fixed
stage parameters, one program produces pools of every sufficiently
large size.

\begin{lemma}
\label{lem:pools}
For every sequence of total computable scan rules \((\pi_i)_{i\ge1}\),
there are Borel sets \(H_i\) in transcript space with complements
\(L_i\) having the following property. Fix \(\ell\ge1\), a rational
\(\eta>0\), and a position bound \(z\). There exist a rational
\(0<\tau<1\) and a program taking a positive integer \(M\) as input.
For every sufficiently large \(M\), it outputs a set \(G\) of \(M\)
distinct positions at least \(z\) satisfying, for every \(i\le\ell\),
\begin{align}
 \kappa\{Z\in H_i:\pi_i\text{ misses a position of }G\}&<\eta,
 \label{eq:pool-high}\\
 \kappa\{Z\in L_i:\pi_i\text{ misses fewer than }\tau M
                  \text{ positions of }G\}&<\eta.
 \label{eq:pool-low}
\end{align}
\end{lemma}

\begin{proof}
For a fair-coin answer sequence \(Z\),
\[
 \int_{2^\N}g_{i,a}(Z)\,d\kappa(Z)
 =\mathbb E_{Z\sim\kappa}[g_{i,a}(Z)]
 =\Pr_{Z\sim\kappa}(\pi_i\text{ ever queries }a).
\]
For a function of the first \(n\) answers, the integral averages its
values over the \(2^n\) prefixes, each with weight \(2^{-n}\).
Giving \(\pi_i\) weight \(w_i=2^{-i}\), define the weighted query
probability and its limit superior by
\[
 p(a)=\sum_{i\ge1}w_i\int g_{i,a}\,d\kappa,\qquad
 p_*=\limsup_{a\to\infty}p(a).
\]
Choose positions tending to infinity along which \(p(a)\to p_*\).
Each integral of \(g_{i,a}\) over a cylinder lies in \([0,1]\).
There are countably many pairs \((i,[v])\), so a diagonal subsequence
\(a_1<a_2<\cdots\) makes \(\int_{[v]}g_{i,a_k}\,d\kappa\) converge
for every \(i\) and every cylinder \([v]\). We use these limits to
construct a function \(g_i\) for each rule. Write \(L^2(\kappa)\)
for the space of measurable functions \(\varphi\) with
\(\|\varphi\|_2^2=\mathbb E_{Z\sim\kappa}[|\varphi(Z)|^2]<\infty\),
identifying functions that agree almost everywhere.

Fix \(i\). For \(Z\) with length-\(d\) prefix \(v\), let \(F_d(Z)\)
be the limiting average on \([v]\),
\[
 F_d(Z)=\frac{1}{\kappa([v])}\lim_{k\to\infty}
        \int_{[v]}g_{i,a_k}(Z')\,d\kappa(Z').
\]
Each \(F_d\) takes values in \([0,1]\). Its value on \([v]\) is the
average of the values of \(F_{d+1}\) on \([v0]\) and \([v1]\).
Write \(\Delta_d=F_{d+1}-F_d\). On each cylinder of depth \(d\),
\(\Delta_d\) has mean zero, while \(F_d\) and all earlier increments
are constant. Thus \(\Delta_d\) is orthogonal to \(F_d\) and to every
earlier increment. For \(m<n\), the Pythagorean identity gives
\[
 \|F_n-F_m\|_2^2
 =\sum_{d=m}^{n-1}\|\Delta_d\|_2^2
 =\|F_n\|_2^2-\|F_m\|_2^2.
\]
The squared norms increase to a limit at most \(1\), since
\(0\le F_d\le1\). Hence \((F_d)\) is Cauchy in \(L^2(\kappa)\), with
limit \(g_i\). A subsequence converges almost everywhere. Its pointwise
limit, set to zero wherever it does not exist, gives a Borel
representative of \(g_i\) taking values in \([0,1]\).
Fix a cylinder \([v]\). For \(d\ge |v|\), summing over the depth-\(d\)
cylinders contained in \([v]\) gives
\[
 \int_{[v]}F_d\,d\kappa
 =\lim_{k\to\infty}\int_{[v]}g_{i,a_k}\,d\kappa.
\]
As \(d\to\infty\), the left-hand side tends to
\(\int_{[v]}g_i\,d\kappa\) by \(L^2\) convergence. Thus the integrals
of \(g_{i,a_k}\) converge to those of \(g_i\) against every cylinder
indicator and, by linearity, against every finite linear combination of
these indicators. Such combinations are dense in \(L^2(\kappa)\).
Since \(\|g_{i,a_k}\|_2,\|g_i\|_2\le1\), it follows that
\(\int\varphi g_{i,a_k}\,d\kappa\to\int\varphi g_i\,d\kappa\)
for every \(\varphi\in L^2(\kappa)\). This is weak convergence in
\(L^2(\kappa)\).

Set \(H_i=\{Z\in 2^\N:g_i(Z)=1\}\) and \(L_i=2^\N\setminus H_i\).
Taking \(\varphi=\mathbf{1}_{H_i}\), the indicator of \(H_i\), gives
\[
 \kappa\{Z\in H_i:g_{i,a_k}(Z)=0\}
 =\int_{H_i}(1-g_{i,a_k}(Z))\,d\kappa(Z)
 \longrightarrow \int_{H_i}(1-g_i(Z))\,d\kappa(Z)=0 .
\]
For any fixed \(M\), we can therefore select \(M\) positions far enough
along the subsequence that the measure of transcripts in \(H_i\) on
which the rule misses even one of them is as small as desired.

Fix \(\ell,\eta,z\). We may decrease \(\eta\) so that \(0<\eta<1\).
By continuity of measure, choose a rational \(0<\tau<1/4\) so that
\[
 \kappa\{Z:1-2\tau<g_i(Z)<1\}<\eta/3 \qquad(i\le\ell).
\]
Choose a rational \(\delta>0\) with \(4\delta<\eta\) and
\(8\delta/\tau<\eta/3\). For each \(i\le\ell\), choose a
clopen set \(H'_i\) and a dyadic rational cylinder function
\(\psi_i\) such that
\[
 \kappa(H_i\mathbin{\triangle}H'_i)<\delta,\qquad
 \|\psi_i-g_i\|_1<\delta.
\]
These approximations exist because the cylinder algebra generates
the Borel sets. Put \(s=w_\ell\delta\). Choose \(z'\ge z\) with
\(p(a)<p_*+s\) for all \(a\ge z'\), a rational
\(\gamma\in(p_*-3s,p_*-2s)\), and an integer \(D\ge\ell\) with
\(\sum_{i>D}w_i<s/4\). The finite data \(z',\gamma,D,\delta\),
the clopen and cylinder approximations, and codes for the first
\(D\) scan rules suffice to specify the program.

For a candidate \(M\)-element set \(G\subseteq\{z',z'+1,\ldots\}\),
let \(v_{i,V}(Z)\) be the fraction of \(G\) that \(\pi_i\) reads in
its first \(V\) queries with answers \(Z\). As \(V\to\infty\), this
fraction increases to \(v_i(Z)=M^{-1}\sum_{a\in G}g_{i,a}(Z)\).
On input \(M\), the program enumerates pairs \((G,V)\). It outputs
\(G\) from the first pair for which the first two inequalities below
both hold for every \(i=1,\ldots,\ell\) and the third holds for the
weighted sum over \(i=1,\ldots,D\).
\begin{align*}
 \kappa\{Z\in H'_i:v_{i,V}(Z)<1\}&<3\delta &&(i\le\ell),\\
 \|v_{i,V}-\psi_i\|_1&<3\delta &&(i\le\ell),\\
 \sum_{i\le D}w_i\int v_{i,V}\,d\kappa&>\gamma.
\end{align*}
All three comparisons involve rational quantities computable from
finitely many transcripts. Later queries may improve coverage on
\(H'_i\) but spoil the approximation to \(\psi_i\). The third test
limits this change. The final weighted average is at most
\(M^{-1}\sum_{a\in G}p(a)<p_*+s\), so after time \(V\) it can
increase by less than \(p_*+s-\gamma\).

To see that the search terminates, use the subsequence \(a_k\).
Choose \(M\)
distinct positions from the subsequence. At each selection require \(a\ge z'\),
\(p(a)>p_*-s/4\), and
\(\kappa\{Z\in H_i:g_{i,a}(Z)=0\}<\delta/M\) for all \(i\le\ell\).
For every position \(a'\) already selected, also require
\[
 \bigl|\langle g_{i,a}-g_i,g_{i,a'}-g_i\rangle\bigr|<\delta^2/4
 \qquad(i\le\ell).
\]
Each step imposes finitely many conditions that hold sufficiently
far along the subsequence. For the resulting set \(G\), expanding
the squared norm and using \(\|g_{i,a}-g_i\|_2\le1\)
gives
\[
 \|v_i-g_i\|_2^2\le 1/M+\delta^2/4 .
\]
Thus \(\|v_i-g_i\|_1<\delta\) for sufficiently large \(M\).
A union bound gives \(\kappa\{Z\in H_i:v_i(Z)<1\}<\delta\).
Averaging \(p(a)\) over \(G\) and removing the tail of the weights
gives \(\sum_{i\le D}w_i\int v_i\,d\kappa>p_*-s/2\).
For this \(G\), \(v_{i,V}\uparrow v_i\) also gives convergence in
\(L^1(\kappa)\), and \(\{v_{i,V}=1\}\uparrow\{v_i=1\}\) since \(G\)
is finite. The strict bounds just obtained, together with the
approximations to \(H_i\) and \(g_i\), give all three search
inequalities at some finite \(V\).

Conversely, let \((G,V)\) be any pair accepted by the program. A
position missed forever was still unread at time \(V\). The first test
and \(\kappa(H_i\mathbin{\triangle}H'_i)<\delta\) therefore give
\eqref{eq:pool-high}. For the later queries, the third test gives
\[
 \sum_{i\le D}w_i\int(v_i-v_{i,V})\,d\kappa
 \le \frac1M\sum_{a\in G}p(a)-\sum_{i\le D}w_i\int v_{i,V}\,d\kappa
 <p_*+s-\gamma<4s.
\]
The differences are nonnegative and \(w_i\ge w_\ell\) for
\(i\le\ell\), so \(\|v_i-v_{i,V}\|_1<4\delta\). The second test
and the approximation to \(g_i\) now give \(\|v_i-g_i\|_1<8\delta\).
On \(L_i\), outside a set of measure less than \(\eta/3\), we have
\(g_i\le1-2\tau\). On this part of \(L_i\), reading more than a
\(1-\tau\) fraction of \(G\) forces \(|v_i-g_i|>\tau\), an event
of measure at most \(8\delta/\tau<\eta/3\) by Markov's inequality.
This proves \eqref{eq:pool-low}. All non-effective choices enter
through the finite data fixed before the program receives \(M\).
\end{proof}

\subsection{Sparse supports and rank preservation}
\label{sec:sparse-matrices}

Number the positions in the pool \(G\) from \(1\) to \(M\) in increasing
order. We need a list of pairs \((S,B)\) with two properties. Each
position occurs in only a small fraction of the supports \(S\). Moreover, for each possible set
\(D\) of at least \(\tau M\) unread positions, the columns indexed by
\(S\cap D\) retain full row rank for most pairs. The exceptional pairs
may depend on \(D\).

\begin{lemma}
\label{lem:design}
Fix rational numbers \(0<\theta,\tau<1\). For every sufficiently large
integer \(r\), let \(M=r^8\), \(m=r^2\), and \(J=M^3\). There is a list
of candidate pairs \((S^{(q)},B^{(q)})_{q=1}^J\). Each support
\(S^{(q)}\subseteq\{1,\ldots,M\}\) has size \(m\), and \(B^{(q)}\) is a
binary \(r\)-by-\(m\) matrix of full row rank with columns indexed by
\(S^{(q)}\). The list satisfies both bounds
\begin{align*}
 \frac1J\#\{q:a\in S^{(q)}\}
   &\le \frac{2m}{M}
   &&\text{for every }a\in\{1,\ldots,M\},\\
 \frac1J\#\{q:\operatorname{rank}(B^{(q)}_{S^{(q)}\cap D})<r\}
   &\le\theta
   &&\text{for every }D\subseteq\{1,\ldots,M\}\text{ with }|D|\ge\tau M.
\end{align*}
A finite search computes such a list from \(r,\theta,\tau\) whenever
one exists.
\end{lemma}

\begin{proof}
Sample \(J\) candidate pairs independently. For each pair, choose \(S\)
uniformly among the \(m\)-element subsets of the pool. Index the columns
of \(B\) by \(S\) in increasing order and draw the matrix uniformly
from the binary \(r\)-by-\(m\) matrices of full row rank.
Fix a set \(D\) of at least \(\tau M\) unread positions. For one random
pair \((S,B)\), put \(N=|S\cap D|\) and
\(p_D=\Pr(\operatorname{rank}(B_{S\cap D})<r)\).
Sampling without replacement gives
\(\mu=\mathbb E N=m|D|/M\ge\tau m\) and
\(\operatorname{Var}(N)\le\mu\), since the membership indicators for
distinct draws have nonpositive covariance. For large \(r\) we have
\(\mu\ge4r\), so Chebyshev's inequality gives
\(\Pr(N<2r)\le 4/\mu\le 4/(\tau m)\).

Fix a support \(S\) with \(N\ge2r\). First draw the entries of the
matrix as independent fair bits. Each nonzero row combination vanishes
on the \(N\) columns in \(S\cap D\) with probability \(2^{-N}\).
A union bound over row combinations gives failure probability at most
\((2^r-1)2^{-N}\) on these columns. Applied to all \(m\) columns, the
same bound gives probability at least \(1-(2^r-1)2^{-m}>1/2\) of
full row rank for large \(r\). Conditioning on full row rank gives
the required matrix law and increases the failure bound by at most a
factor of two, to \(2(2^r-1)2^{-N}<2^{1-r}\).
Combining the two cases yields
\[
 p_D\le \Pr(N<2r)+\Pr(N\ge2r,\ \operatorname{rank}(B_{S\cap D})<r)
     \le \frac{4}{\tau m}+2^{1-r}<\frac{\theta}{4}
\]
for all sufficiently large \(r\), uniformly in \(D\).

For the full list, let \(Y_D\) count the candidates that lose rank
on \(D\). Independence gives \(\mathbb E 2^{Y_D}=(1+p_D)^J\le\exp(p_D J)\),
so
\[
 \Pr(Y_D>\theta J)\le \exp\!\bigl[-(\log 2-1/4)\theta J\bigr].
\]
For a fixed position \(a\), the number \(\nu(a)=\#\{q:a\in S^{(q)}\}\)
is binomial with parameters \(J\) and \(m/M\). The same moment bound gives
\[
 \Pr(\nu(a)>2mJ/M)\le \exp\!\bigl[-(2\log 2-1)mJ/M\bigr].
\]
There are at most \(2^M\) sets \(D\) and \(M\) positions \(a\).
Since \(J=M^3\) and \(mJ/M=mM^2\), summing the two failure bounds
over these choices gives a quantity tending to zero. Thus a list
satisfying both requirements exists for large \(r\). Both requirements
are decidable on a finite list, so exhaustive search finds one.
\end{proof}

\section{Choosing the affine constraints}
\label{sec:construction}

At stage \(j\) we choose one parity system. Its solution set must enter
level \(U_j\) of the Martin-L\"of test, and its contribution to the
capital estimates for the first \(j\) strategies must be small enough
to sum over \(j\).

For this choice, Lemma~\ref{lem:pools} supplies a pool that each strategy,
given fair-coin answers, is likely either to read in full or to leave
a fixed fraction unread. In the latter case, Lemma~\ref{lem:design}
ensures that for most candidate pairs \((S,B)\) the unread columns
retain full rank. Reading the entire pool, on the other hand, forces
rank drops on every candidate block. To bound the wagers, we use both
the squared-sum bound of Lemma~\ref{lem:budget} and the small fraction of
candidate supports containing any given position (Lemma~\ref{lem:design}).
Together these give a small total wager on \(S\), averaged over
candidates. We sum the wager and rank-failure estimates over the first
\(j\) strategies before choosing a pair, so the same choice serves all
of them.

Fix a list of all total computable strategies and normalize each one by
Lemma~\ref{lem:budget}. The list need not be effective. For its \(i\)th
entry, write \(\pi_i\) for the scan rule and \(h_i\) for the signed
wager fraction of the normalized strategy.
Apply Lemma~\ref{lem:pools} to the scan rules to obtain \(H_i,L_i\).
For an input law \(P\), we write \(P(F)\) for the probability that
the transcript under \(\pi_i\) belongs to \(F\) when the rule is clear
from the indices. In particular, the transcript cylinder \([w]\) has
probability \(P(C_{\pi_i}(w))\).

For each sufficiently large trial rank \(r\), we choose, in order, a
support \(S_j\) disjoint from the earlier blocks, a full-row-rank
matrix \(B_j\) on \(S_j\) with \(r\) rows, a query horizon \(V_j\),
and a syndrome \(y_j\in\{0,1\}^r\). We can therefore take the final
rank \(r_j\) large enough to meet the description bound
\eqref{eq:test-criterion}, as we verify below. Write
\[
 E_j=\{X:B_j(X|_{S_j})=y_j\},\qquad
 P_j=\lambda(\,\cdot\mid E_1\cap\cdots\cap E_j),\qquad P_0=\lambda .
\]

For the scan rule \(\pi_i\), let \(R_{ij}(t)\) be the \emph{remaining rank}
on block \(j\) after \(t\) queries, namely the rank of the unread
columns of \(B_j\). Initially \(R_{ij}(0)=r_j\). The first query that
lowers this rank has index
\[
 T_{ij}=\min\{t\ge1:R_{ij}(t)<R_{ij}(t-1)\},
 \qquad \min\varnothing=\infty.
\]
The remaining rank is \emph{partial} when
\(0<R_{ij}(t)<r_j\). Let \(W_{ij}(t)\) be the sum of the fractions
of capital wagered, \(|h_i|\), at rank drops on block \(j\) through
query \(t\). Each fraction is evaluated before the answer is revealed.

Later supports avoid every position that any of the first \(j\) rules
can query in its first \(V_j\) steps, on any transcript. Later
constraints thus preserve the laws of these transcripts up to that
horizon. Put \(e_j=2^{-j}/100\). For every \(i\le j\) we require
\begin{align}
 \mathbb E_{P_j} W_{ij}(V_j) + P_j(0<R_{ij}(V_j)<r_j)
     &<4e_j, \label{eq:short-cost}\\
 P_j(V_j<T_{ij}<\infty)
     &\le 2e_j, \label{eq:first-late}\\
 P_{j-1}(H_i\cap\{T_{ij}=\infty\})
     &<e_j, \label{eq:high-cover}\\
 P_a(L_i\cap\{T_{ij}<\infty\})
     &<e_j \qquad(0\le a<j). \label{eq:low-ignore}
\end{align}

The first two bounds distinguish three possibilities at \(V_j\). At
rank zero there can be no later drop on block \(j\). Partial rank is
covered by \eqref{eq:short-cost}, together with the wagers already made
at drops. At full rank any later drop must be the first and is covered
by \eqref{eq:first-late}. The last two bounds concern the probabilities
of transcripts in \(H_i\) with no drop on block \(j\) and of transcripts
in \(L_i\) with a drop there. We
need \eqref{eq:low-ignore} under each earlier law \(P_a\). For
\(1\le a<j\), the first drop on any block \(k\ge a\) after \(V_a\)
can occur on block \(j\).

\paragraph{Support and matrix.}
Fix the choices at earlier stages and put \(K=\sum_{a<j}r_a\). Disjointness of
the supports gives \(\lambda(E_1\cap\cdots\cap E_a)=2^{-\sum_{k\le a}r_k}\).
Thus \(P_a\) has density at most \(2^K\) relative to \(\lambda\) for
every \(0\le a<j\). For a fixed scan rule, taking the input preimage
of a transcript event \(F\) gives \(P_a(F)\le 2^K\kappa(F)\), since
fair-coin inputs induce the transcript law \(\kappa\).
Choose \(z\) above all earlier supports and,
for each \(1\le a<j\), above every position any of the first \(a\) scan
rules can query through \(V_a\). Only finitely many positions are
excluded, since each finite horizon admits finitely many transcripts.

Apply Lemma~\ref{lem:pools} with \(\ell=j\), this bound \(z\), and
\(\eta=e_j/(100j2^K)\). Fix the resulting pool program and omission
fraction \(\tau\). For every sufficiently large trial rank \(r\), the
program returns a pool \(G\) of \(M=r^8\) positions. Identify the
column indices \(1,\ldots,M\) with the positions of \(G\) in increasing order,
and compute the list of Lemma~\ref{lem:design} with \(\theta=\eta\).
Write the resulting list as \((S^{(q)},B^{(q)})_{q=1}^J\), where
\(J=M^3\) and each support has size \(m=r^2\).

Fix \(i\le j\). Every candidate support \(S\) lies in \(G\), so the
\(\kappa\)-measure of transcripts in \(H_i\) on which the rule misses
part of \(S\) is less than \(\eta\) by \eqref{eq:pool-high}.
On \(L_i\), outside a set of \(\kappa\)-measure less than \(\eta\),
the rule leaves a set \(D\) of at least \(\tau M\) pool positions
unread by \eqref{eq:pool-low}. For a fixed transcript this is the same
\(D\) for every candidate. All but at most \(\theta J\) pairs have
\(B_{S\cap D}\) of rank \(r\), and for these pairs the remaining rank
stays full throughout the run. Averaging over the list, the
\(\kappa\)-probability of a transcript in \(L_i\) with a rank drop
is therefore at most \(\eta+\theta\).

For any finite set \(S\) of positions, let \(A_i(S)\) be the sum of
the wager fractions \(|h_i(v_t)|\) over those queries for which
\(\pi_i(v_t)\in S\). Each pool position
is queried at most once. By Cauchy--Schwarz and \eqref{eq:budget},
\(A_i(G)\le\sqrt{M/2}\) on every transcript. Each position lies in at
most a fraction \(2m/M\) of the supports, so the average of
\(\mathbb E_{P_{j-1}}A_i(S)\) over the list is at most
\[
 \frac{2m}{M}\sqrt{M/2}=\sqrt{2}\,r^{-2}.
\]
For each candidate pair \((S,B)\), let \(T_i(S,B)\) be the first query
by \(\pi_i\) that lowers the rank of the unread columns of \(B\),
with \(T_i(S,B)=\infty\) if no such query occurs.
For each \(i\le j\), the two bounds above are averages over the same
candidate list. Multiplying the rank-drop bound by \(2^K\), adding the
wager bound, and summing over \(i\le j\) gives
\[
 \frac1J\sum_{q=1}^J\sum_{i\le j}
 \left(\mathbb E_{P_{j-1}}A_i(S^{(q)})
       +2^K\kappa(L_i\cap\{T_i(S^{(q)},B^{(q)})<\infty\})\right)
 \le \sqrt{2}\,j r^{-2}+2^Kj(\eta+\theta).
\]
The right-hand side is less than \(e_j\) for all sufficiently large
\(r\). Choose an index \(q\) for which the inner sum is less than \(e_j\)
and write \((S_j,B_j)=(S^{(q)},B^{(q)})\) and \(r_j=r\) for this trial.
For this choice,
\(T_{ij}=T_i(S_j,B_j)\). By the density bound, for each
\(a<j\) the left-hand probabilities in \eqref{eq:low-ignore} sum to
less than \(e_j\) over \(i\le j\). Likewise, the sum over \(i\le j\)
of the \(P_{j-1}\)-probabilities of a transcript in \(H_i\) missing part
of \(S_j\) is less than \(2^Kj\eta<e_j\). If \(\pi_i\) reads all of
\(S_j\), the remaining rank falls from \(r_j\) to zero and hence drops
at least once. This proves \eqref{eq:high-cover}.

\paragraph{Horizon and syndrome.}
For \(t\ge0\), put
\[
 u(t)=\sum_{i\le j}P_{j-1}(0<R_{ij}(t)<r_j),\qquad
 v(t)=\sum_{i\le j}P_{j-1}(T_{ij}\le t).
\]
Under \(P_{j-1}\), \(u(t)\) is the expected number of strategies
among the first \(j\) with partial rank on this block, and \(v(t)\)
is the expected number that have caused a drop by time \(t\).
Both are computable rationals. For fixed \(t\) we enumerate the
length-\(t\) transcripts and compute \(P_{j-1}(C_{\pi_i}(w))\)
by counting compatible assignments on the queried positions and the
earlier blocks. Because \(S_j\) is finite, its remaining rank
stabilizes on every transcript. On \(H_i\), partial limiting rank
requires \(\pi_i\) to miss a position of \(S_j\). On \(L_i\), it
requires a rank drop, so \(T_{ij}<\infty\). For the selected pair
\((S_j,B_j)\) we proved
\[
 \sum_{i\le j}P_{j-1}\{Z\in H_i:\pi_i\text{ misses a position of }S_j\}<e_j,
 \qquad
 \sum_{i\le j}P_{j-1}(L_i\cap\{T_{ij}<\infty\})<e_j .
\]
Hence \(\lim_t u(t)<2e_j\) by bounded convergence. Also \(v(t)\) increases
to a limit \(v_\infty\le j\).

To recover a suitable horizon from a short program we need a stopping
rule for \(v(t)\). Give the program the integer
\(b_j=\lfloor v_\infty/e_j\rfloor\) as advice. It takes the first
\(V_j\) for which \(u(V_j)<3e_j\) and \(v(V_j)>(b_j-1)e_j\).
Such a time exists by the two limits above. Since
\(b_je_j\le v_\infty<(b_j+1)e_j\),
\[
 \sum_{i\le j}P_{j-1}(V_j<T_{ij}<\infty)
 = v_\infty-v(V_j)<2e_j .
\]
This estimate also holds after conditioning on any syndrome at stage
\(j\). For the rule \(\pi_i\), the input event \(V_j<T_{ij}<\infty\)
is the disjoint union of the sets \(C_{\pi_i}(w)\) for transcripts
\(w\) immediately before the first drop, with \(|w|\ge V_j\).
The next position, and hence the drop, is determined by \(w\). On
each such input event \(C=C_{\pi_i}(w)\), the unread columns of
\(B_j\) still have rank \(r_j\). If \(P_{j-1}(C)>0\), the unread bits
in \(S_j\) are independent fair bits conditional on \(C\), since \(S_j\)
is disjoint from the earlier blocks. Thus for every syndrome \(y\),
\[
 P_{j-1}\bigl(C\cap\{B_j(X|_{S_j})=y\}\bigr)=2^{-r_j}P_{j-1}(C).
\]
The identity also holds when \(P_{j-1}(C)=0\). Summing over the
disjoint input events and dividing by \(2^{-r_j}\) proves
\eqref{eq:first-late} for every choice of \(y_j\).

For each syndrome \(y\), sum the left side of \eqref{eq:short-cost}
over \(i\le j\) under \(P_{j-1}\) conditioned on \(B_j(X|_{S_j})=y\).
This sum is computable by enumerating length-\(V_j\) transcripts, since
the wager fractions and the conditional cylinder probabilities are
rational. Averaging over \(y\) gives the same sum under \(P_{j-1}\),
where all syndromes are equally likely. Since
\(W_{ij}(V_j)\le A_i(S_j)\) on every transcript, the wager estimate
for the selected candidate gives
\[
 \sum_{i\le j}\mathbb E_{P_{j-1}}W_{ij}(V_j)
 \le \sum_{i\le j}\mathbb E_{P_{j-1}}A_i(S_j)<e_j .
\]
Partial rank contributes \(u(V_j)<3e_j\). The average is thus less than \(4e_j\).
Take \(y_j\) to be the lexicographically first minimizing syndrome.
Its sum is less than \(4e_j\), giving \eqref{eq:short-cost} for each
\(i\le j\).

\paragraph{Description length.}
For the interpreter fixed in Section~\ref{sec:test}, encode the earlier
affine constraints, programs computing \(\pi_i\) and \(h_i\) for
\(i\le j\), the pool program, and \(j,z,\eta,\tau,e_j\) in a finite header.
The header is independent of the trial rank \(r\). Given \(r\), the
program reconstructs the candidate list. To specify the chosen pair
it needs only the index \(q\) supplied by the averaging argument.
With \(b_j\) it also recovers \(V_j\) and computes the first minimizing
syndrome. Encoding \(y_j\) directly would cost \(r\) bits.

The list has \(J=r^{24}\) entries, so the index and \(r\) together
require \(O(\log(r+2))\) bits. Although \(b_j\) can depend on the
trial rank, \(0\le b_j\le j/e_j\), a bound fixed in advance. Encoding
the data in self-delimiting fields thus gives a plain program \(p_j\)
with
\[
 |p_j|\le C_j+O(\log(r+2)),
\]
where \(C_j\) is independent of \(r\). Suitable choices exist for
every sufficiently large \(r\), so we may choose \(r_j\) large enough
that \(r_j\ge 2|p_j|+j+1\). Each affine constraint has this description
bound even though the sequence of constraints need not be computable.

\section{Bounded capital and proof of the separation}
\label{sec:capital}

Let \(\rho\) be the product law obtained by sampling a uniform solution
on each block and a fair bit at each remaining position. Its support
is \(\bigcap_j E_j\). Indeed, this set is closed and has full measure,
and any cylinder meeting it specifies compatible values on finitely
many blocks and free positions, an event of positive measure. By
\eqref{eq:test-criterion} and the description bound from
Section~\ref{sec:construction}, every sequence in the support fails
the test \((U_m)_{m\ge1}\).

Fix a strategy \(i\). The bounds of Section~\ref{sec:construction}
were proved under the finite-stage laws \(P_j\). To use them under
\(\rho\), for each \(j\ge i\) we compare transcript probabilities
through \(V_j\) and just before the first subsequent rank drop on
any block \(k\ge j\).

\paragraph{Comparing transcript probabilities.}
Fix a transcript \(w\) of \(\pi_i\). Its finite assignment meets only
finitely many blocks. On any block \(k\) whose unread columns retain
full rank, the syndrome is still uniform under fair-coin sampling
given the assignment. Conditioning on that syndrome leaves the
probability of the assignment unchanged (Section~\ref{sec:test}).
Independence across blocks therefore gives
\(\rho(C_{\pi_i}(w))=P_j(C_{\pi_i}(w))\) whenever every block \(k>j\)
still has full rank after \(w\). In particular, for \(j\ge i\) the
two laws agree through \(V_j\), since \(\pi_i\) cannot have queried
any later block by then.

We also compare the transcript laws induced by \(P_j\) and \(P_{j-1}\)
on the event \(T_{ij}=\infty\). For a transcript cylinder \([w]\)
and \(n\ge |w|\), the set \([w]\cap\{T_{ij}>n\}\) is a finite union
of length-\(n\) transcript cylinders. Their probabilities agree under
the two laws because the remaining rank on block \(j\) is still full.
Taking the decreasing limit gives equality on \([w]\cap\{T_{ij}=\infty\}\).
Since the cylinders generate the Borel sets, for every Borel transcript
event \(F\) we obtain
\begin{equation}
 P_j(F\cap\{T_{ij}=\infty\})
 =P_{j-1}(F\cap\{T_{ij}=\infty\}).
 \label{eq:restricted}
\end{equation}

\paragraph{Total wager at rank drops.}
Fix \(j\ge i\). If \(R_{ij}(V_j)=0\), no later query can cause a
rank drop on block \(j\). Partial rank at \(V_j\) has \(\rho\)-probability
at most \(4e_j\) by \eqref{eq:short-cost}, since \(\rho\) and \(P_j\)
agree on the first \(V_j\) answers. It remains to consider full rank.
If we wait for the first drop on block \(j\), some block \(k>j\) may
already have had a drop, and the two laws may then assign different
probabilities to the transcript. We therefore consider the first drop
after \(V_j\) on \emph{any} block \(k\ge j\).

Let \(\mathcal A_{ij}\) be the event that block \(j\) has full rank
at \(V_j\) and a drop subsequently occurs on some block \(k\ge j\).
Take the transcript \(w\) immediately before the first such drop.
It already determines the next position and whether its query lowers
the rank, so the sets \(C_{\pi_i}(w)\) form a disjoint cover of
\(\mathcal A_{ij}\). No block \(k>j\) was queried by \(V_j\), and
none has lost rank by the end of \(w\). Thus
\(\rho(C_{\pi_i}(w))=P_j(C_{\pi_i}(w))\) for each such transcript.
Summing over these disjoint events gives
\(\rho(\mathcal A_{ij})=P_j(\mathcal A_{ij})\).

On \(\mathcal A_{ij}\cap\{T_{ij}<\infty\}\), the first drop on block \(j\)
is later than \(V_j\), an event of \(P_j\)-probability at most
\(2e_j\) by~\eqref{eq:first-late}. On \(\mathcal A_{ij}\cap\{T_{ij}=\infty\}\),
the inputs whose transcripts lie in \(H_i\) have \(P_j\)-probability
less than \(e_j\) by~\eqref{eq:restricted} with \(F=H_i\) and
by~\eqref{eq:high-cover}.
Each remaining input in \(\mathcal A_{ij}\) has a transcript in \(L_i\) and a
rank drop on some later block \(k>j\). For each \(k\), the stage-\(k\)
estimate~\eqref{eq:low-ignore} applies to \(P_j\). A union bound gives
\[
 \rho(\mathcal A_{ij})\le 3e_j+\sum_{k>j}e_k .
\]
A late drop on block \(j\) from full rank at \(V_j\) is included in
\(\mathcal A_{ij}\). Adding the partial-rank bound and using
\(e_j=2^{-j}/100\) gives
\[
 \rho(\text{a rank drop on block \(j\) after \(V_j\)})
 \le 7e_j+\sum_{k>j}e_k=8e_j.
\]
By the first Borel--Cantelli lemma, almost surely only finitely many
blocks \(j\ge i\) have a drop after \(V_j\). Each of these, and each
of the finitely many blocks \(j<i\), contributes at most \(r_j\)
drops. For the drops through \(V_j\) on blocks \(j\ge i\), we instead
sum the wagers. Agreement of \(\rho\) and \(P_j\) through \(V_j\) and
\eqref{eq:short-cost} give
\[
 \mathbb E_\rho\sum_{j\ge i}W_{ij}(V_j)\le 4\sum_{j\ge i}e_j<\infty .
\]

Combining the wagers through the horizons with those at the finitely
many remaining drops gives
\begin{equation}
 \sum_{\substack{t\ge0\\ \text{query }t+1\text{ causes a rank drop}}}
 |h_i(v_t)|<\infty
 \qquad \rho\text{-almost surely}.
 \label{eq:total-drop-wagers}
\end{equation}

\paragraph{Capital of the normalized strategy.}

At every query with no rank drop, either the bit lies outside all
blocks or deleting its column preserves the span of the unread columns
in its block. The conditional law given the transcript is fair in
both cases by independence across blocks and the calculation in
Section~\ref{sec:test}.

Let \(I_t\) indicate that query \(t+1\) causes no rank drop. The
transcript \(v_t\) determines \(I_t\) and \(h_i(v_t)\). When \(I_t=1\),
\(b_t\) is conditionally fair, so
\[
 \mathbb E_\rho\!\left[I_t(2b_t-1)h_i(v_t)\mid v_t\right]=0 .
\]
Consequently
\[
 M_n^{\mathrm{fair}}=\sum_{t<n} I_t (2b_t-1)h_i(v_t)
\]
is a \(\rho\)-martingale. Orthogonality of its increments and
Lemma~\ref{lem:budget} give
\[
 \mathbb E_\rho (M_n^{\mathrm{fair}})^2
 =\mathbb E_\rho\sum_{t<n}I_t h_i(v_t)^2\le \tfrac12.
\]
The \(L^2\) martingale convergence theorem implies that \(M_n^{\mathrm{fair}}\)
converges almost surely. Let \(D_n\) be the normalized strategy's
capital after \(n\) queries, as in Lemma~\ref{lem:budget}. Applying
\(\log(1+x)\le x\) to each capital update gives
\[
 \log D_n\le M_n^{\mathrm{fair}}+\sum_{t<n}(1-I_t)|h_i(v_t)|.
\]

The first term converges almost surely. The second term is a partial
sum of the nonnegative series in \eqref{eq:total-drop-wagers}, so it
is almost surely bounded uniformly in \(n\). Thus \(D_n\) is bounded
\(\rho\)-almost surely. By Lemma~\ref{lem:budget}, the original strategy
can succeed only where the normalized strategy succeeds. Taking the
union of these null success sets over the countably many computable
strategies gives Kolmogorov--Loveland randomness \(\rho\)-almost surely.
Together with the support inclusion above, this proves
Theorem~\ref{thm:main}.

\paragraph{Remark.} One can also single out a separating sequence.
For each original strategy \(i\),
choose an integer \(c_i\) such that the probability under \(\rho\) of
ever exceeding capital \(c_i\) is less than \(2^{-i-1}\). Such a bound
exists by almost-sure boundedness. Exceeding \(c_i\) is witnessed by a
finite transcript, so the set on which all bounds hold is closed.
Its intersection with the closed set \(\bigcap_j E_j\) has
\(\rho\)-measure at least \(1-\sum_{i\ge1}2^{-i-1}=1/2\). As a
nonempty closed subset of Cantor space it has a lexicographically
least member, which is a separating sequence.

\section*{Use of generative AI}
We used OpenAI's GPT-6.2 Astra to help construct the probability measure
and develop the proof of Theorem~\ref{thm:main}. We also used it in
drafting and revising the manuscript. The authors take responsibility
for its contents.

\end{document}